\documentclass{article}

\usepackage{graphicx}
\usepackage{subfig}

\begin{document}

\title{Novel Side Channel Attacks in Continuous Variable Quantum Key Distribution}

\author{Elizabeth Newton\thanks{email: py10en@leeds.ac.uk \newline School of Physics and Astronomy, University of Leeds, Leeds, LS2 9JT}         \and
        Matthew C.J. Everitt	\and
        Freya L. Wilson	\and
        Benjamin T.H. Varcoe }

\date{18/06/2015}

\maketitle

\begin{abstract}
Experimental Quantum Key Distribution (QKD) protocols have to consist of not only the unconditionally secure quantum transmission, but also a subsequent classical exchange that enables key reconciliation and error correction. There is a large body of work examining quantum attacks on the quantum channel, but here we begin to examine classical attacks to both the classical communication and the exchange as a whole. Linking together separate secure protocols can unexpectedly leak information to an eavesdropper, even if the components are unconditionally secure in isolation. Here we focus specifically on the join between quantum and classical protocols, finding that in just this crossing of the quantum-classical boundary, some security is always and unintuitively lost. This occurs with no communication between the separate parties. While this particular example applies to only Continuous Variable Quantum Key Distribution (CVQKD), it highlights the need to re-examine the way all individual protocols are actually used.

\end{abstract}

\section{Introduction}
Quantum Key Distribution (QKD) \cite{rev} has the potential to create completely secret communications, and has therefore predictably been received with interest by industrial and security sectors \cite{toshiba}\cite{Munro}. It is now a mature technology, with commercial QKD systems already available, meaning that evaluating the practical security of real QKD systems has become essential. 

\begin{figure}
\begin{center}
\includegraphics[width=\linewidth]{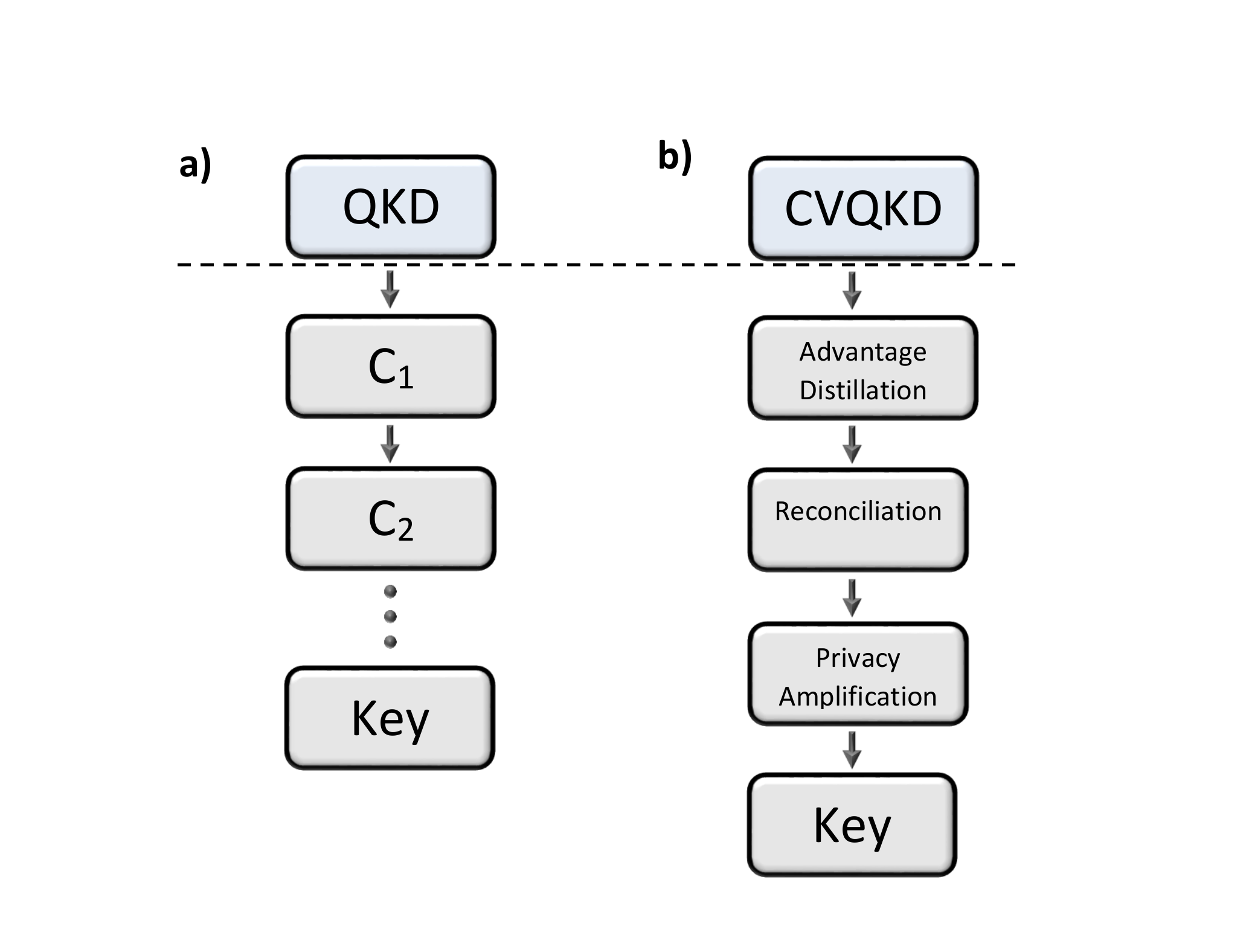}
\caption{{\bf a$)$} The different stages of a QKD protocol. The dashed line separates the quantum transmission from the subsequent classical stages, or blocks ($C_1, C_2$) such as reconciliation, privacy amplification or error correction. The dots indicate the possibility of subsequent classical blocks between the ones depicted and the final key, dependant on the particular protocol used. {\bf b$)$} This is an example showing the classical bocks typically used during a CVQKD protocol, after the continuous variable quantum transmission.}
\label{fig:1}
\end{center}
\end{figure}

While it has been proved that a quantum transmission can be unconditionally secure \cite{rev}, in figure \ref{fig:1} we can see that in a real system the quantum transmission only makes up a small part of the whole QKD protocol. The quantum part is invariably followed by classical communications steps, usually at least one of key reconciliation, privacy amplification or error correction.

In protocols such as BB84, and in all implementations of CVQKD, the classical component is an essential part of the protocol. In others, it at least has to exist as a form of practical error correction to eliminate experimental errors. It is currently not, and may never be, possible to prevent these experimental errors, implying that currently, if not permanently, a classical communication step is unavoidable. 

Proofs of quantum security do not take into account side channel attacks on either the quantum or the classical channel. There has been a lot of work looking at side channel attacks on the quantum transition [4-8]. This work has naturally led to the development of device independent protocols [9-11] which eliminate the risk of side channel attacks to the quantum channel. 

Unfortunately while these device independent protocols protect the quantum transition, the classical channel still remains vulnerable to side channel attacks. Incautious use of the classical component can either reduce the overall security or inadvertently leak information to an eavesdropper. Proofs of quantum security, including device independent protocols, do not consider implementation weaknesses in the classical parts of QKD protocols, leaving even the best quantum protocol open to classical side channel attacks. It is essential to be aware of these potential weaknesses.

\begin{figure}[pb]
\begin{center}
\includegraphics[width=\linewidth]{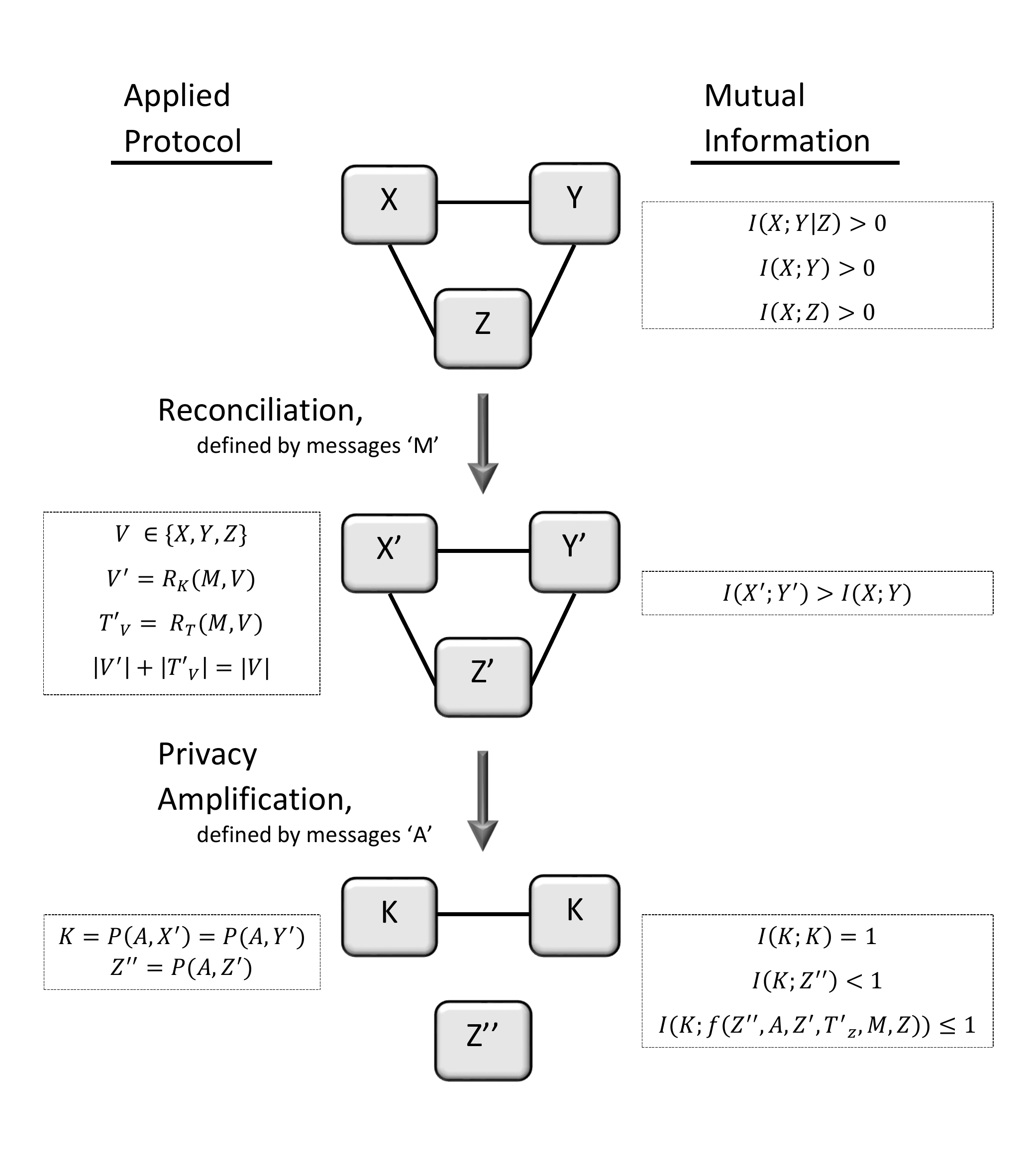}
\caption{An overview of the classical protocols commonly used in CVQKD. After the quantum communication, all three parties know some information about the transmission. Alice's information is denoted $X$, Bob's $Y$ and Eve's $Z$. At this stage, $I(X;Y|Z)>0, I(X;Y)>0$, and $I(X;Z)>0$. During the reconciliation stage, $R$, over a set of messages, $M$, the information known by each party is condensed. If $V \in \{X,Y,Z\}$, then $V' = R_K(M,V)$ to form a new set of key elements, $X'$,$Y'$,$Z'$. The aim is for $I(X';Y')>I(X;Y)$, and for this $|V'|<|V|$. This means there is some information $T'_V$, where $T'_V = R_T(M,V)$, which is left over. Alice and Bob bin this information, but Eve will keep $T'_Z$, as it can be used later to discover more about the system.
After privacy amplification, $P$, defined by a set of messages, $A$, Alice and Bob have managed to agree a key, $K$. Where $K = P(A,X') = P(A, Y')$. Eve is left with $Z'' = P(A, Z')$ where $I(K,Z'')<1$. This is the condition for secrecy. 
Unfortunately, Eve has much more information available to her than just $Z''$. During every stage of the protocol, she gains some information about the system. If instead of throwing this information away, as Alice and Bob do, she keeps it, then she can construct a function to use this and any information she gained during the quantum transmission to cascade back through the different blocks and gain more information about the key. The protocol is in fact only secure if $I(K;f(Z,Z'',M,Z'))<1$.}
\label{fig:2}
\end{center}
\end{figure}

In CVQKD, the quantum exchange must be followed by at least two different classical protocols, key reconciliation and privacy amplification. Figure \ref{fig:2} presents a typical sequence for arriving at a key during CVQKD. While these separate protocol `blocks' can each be proven to be individually secure, very little thought is given to the security of a combination of multiple blocks run in sequence. The problem with chaining together multiple blocks is that information obtained by an eavesdropper (Eve) from each block could be cascaded to reveal more information about the key.

For the sequence in figure \ref{fig:2}, which follows the quantum exchange, the two legitimate users (Alice and Bob) each have some information about the quantum transmission. The information known by Alice is denoted X and, that by Bob, Y. It is also possible that Eve will have gained some knowledge of the transmission, and this is denoted Z. In information theory, information overlap and secrecy are measured using the mutual information, $I(X;Y)$, and conditional information, $I(X;Y|Z)$, respectively \cite{Maurer}. These are defined to be:
\begin{equation}
I(X;Y) = \sum_{y \in Y}^{} \sum_{x \in X}^{} p(x,y) \log_2 \left(\frac{p(x,y)}{p(x)p(y)}\right)
\end{equation} 
\begin{equation}
I(X;Y|Z) = \sum_{z \in Z}^{} \sum_{y \in Y}^{} \sum_{x \in X}^{} p(x,y,z) \log_2 \left(\frac{p(z)p(x,y,z)}{p(x,z)p(y,z)}\right)
\end{equation}

Where $p(x)$ and $p(y)$ are the marginal probability distribution functions of $X$ and $Y$ respectively, and $p(x,y)$ is the joint probability distribution function of $X$ and $Y$.
For a secure channel, after the quantum exchange, the following are in general true:

\begin{equation}
I(X;Y|Z)>0
\end{equation} 
\begin{equation}
I(X;Y)>0
\end{equation} 
\begin{equation}
I(X;Z)>0
\end{equation} 

During the reconciliation step, $R$, defined by a set of messages `$M$', the information known by Alice is reconciled with that known by Bob to produce new key elements, $X'$ and $Y'$, by $X' = R_K(M,X)$ and similarly for $Y$. However, as $|X'|<|X|$ and $|Y'|<|Y|$, there is some information, $T'_X$ and $T'_Y$, which is thrown away. At this point, $I(X';Y')>I(X;Y)$. The eavesdropper can follow exactly the same process, arriving at $Z' = R_K(M,Z)$. However, unlike Alice and Bob, it would be foolish for her to throw away her excess information, $T'_Z$, as this can be used later to give her more information about the system.

After the subsequent step, privacy amplification, $P$, defined by a set of messages `$A$', Alice and Bob are each left with a key, $K$, where $K = P(A,X') = P(A,Y')$. Eve should emerge from the privacy amplification with $Z'' = P(A,Z')$ where $I(K;Z'') < 1$. This ensures secrecy.
Unfortunately, there is more information available to Eve than just $Z''$. She can compile a function which allows her to extract any excess information revealed by the classical exchange above any information she received quantumly. Although there may be isolated circumstances where Eve receives precisely zero excess information, in general the system is secure if and only if the following holds: 
\begin{equation}
I(K;f(Z'',A,Z',T'_Z,M,Z))<1
\end{equation}
Where $f$ is  the function compiled by Eve to maximise her knowledge of $K$. However note that $f$ may be either unknown or may change from exchange to exchange, nevertheless its existence must be taken into account.

In this paper, we demonstrate that this landscape is more complex than is at first apparent, by showing that simply the transition from a message received quantumly to information processed classically in general lowers the secrecy of CVQKD. This happens even before any classical communication occurs and is the result of local transformations of the data. 

This paper concerns itself solely with this data digitisation step and the importance of this new type of side channel attack, not the subsequent reconciliation. No attempt is made to propose one reconciliation protocol over another.

While this weakness in particular only applies to CVQKD, it highlights the need to consider the classical protocol elements as carefully as the quantum. This case demonstrates a counter intuitive violation of the basic assumption that unbroadcast local operations do not affect the security of the protocol. This violation suggests that other non-quantum protocol components, and the transitions between them, need to be reassessed.

\section{Channel Simulation}
\label{sec:1}
In CVQKD a continuous distribution of numbers is transferred through a quantum channel. It is then transformed into a binary string to form the basis of a secret key. CVQKD was proposed \cite{CVQKDproposed} with the idea of increasing the key rate from that of QKD, whilst also increasing the ease of implementation, and reducing the need for single photon sources and detectors.

In general, in CVQKD [1,13-19] the sender (Alice) applies separate, random Gaussian distributed modulations to the phase and amplitude quadratures of a laser. The receiver (Bob) then measures either (or both \cite{noswitching}) quadratures, obtaining a Gaussian distribution of random numbers. Noise introduced into the system through any of a number of sources such as shot noise, channel noise, detector noise or an eavesdropper, will mean that each of the points Bob measures will have a probability of some error, with respect to that originally sent by Alice. In the case of Gaussian additive noise, from Bob's perspective, the value sent by Alice has a Gaussian probability distribution centred on the value received by Bob. At the end of the process, Alice and Bob are left with non-identical distributions of continuous random numbers.

In order for Bob and Alice to reconcile a key, each of their continuous distributions of numbers have to be converted into a binary string. 
There are a number of different ways in which this can be done. Here, in a bid to aid transparency only one method, called slicing, is examined. Slicing has been succeeded by protocols which allow communication across longer distances, and are more optimal to use in practice, such as the one described by Leverrier et al. in \cite{MDRA}.
This conversion is the first and simplest thing that happens to the data when it exits the quantum channel, and even this has a security risk associated with it. There are a number of different methods of slicing, with different levels of security, ease of implementation and key production rates.

The simplest method of slicing is to take values which fall in the positive side of the Gaussian mean as binary `1', and negative values as binary `0'. It can be easily argued that very few of the errors in transmission will be converted to errors in the bit string. Only those points with an error margin that crosses between the positive and negative sides of the quadrature can create errors in the final bit string. This enables production of one bit per transmitted point, and the errors that do get transferred are later removed using standard classical error correction protocols.

Slicing is purely a local output, however even this has a security risk associated with it. There are two methods of slicing which show the extremes of these security implications. In the first, every bit of data is encoded as a `0'. This has no security, and no ability to communicate. In the second, each bit of data is encoded randomly as either `0' or `1'. This is completely secure, but again, communication is impossible. A realistic slicing method has to find some middle ground between these two cases, where communication is possible, and the security is maximised.

It is possible to obtain a higher information transfer, with the same transmission rate, by dividing (or `slicing') the Gaussian distribution of numbers into a larger number of sections, referred to as `bins'. A change of slicing with two bins (producing one string bit), to slicing with four bins (allowing two string bits for each transmitted point), effectively doubles the data rate. This method does however also increase the number of transmission errors transferred into bit string errors, subsequently referred to as `transferred errors', due to the higher number of boundaries between bins.

In a worst case scenario, an eavesdropper (Eve) will have managed to gain significant information about the quantum transmission. She will have measured a Gaussian distribution of random numbers, different from those of both Bob and Alice. As Alice and Bob can only discuss which slicing method they are going to use over classical channels, to which Eve can listen, Eve will know which method they are using, and can apply it to her own data in an attempt to keep her key as similar as possible to that of Alice and Bob. As slicing transfers transmission errors into bit string errors and the different slicing methods transfer different numbers of errors, the slicing method best to use should be chosen on an analysis of the number of transmission errors between Alice and Bob, and also between Eve and the legitimate users.

\begin{figure}
\begin{center}
\parbox{\textwidth}{\includegraphics[width=\textwidth]{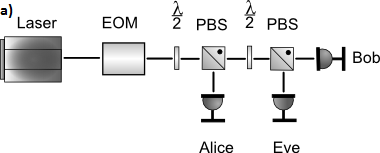}}
\\
\qquad
\begin{minipage}{0.5\textwidth}%
\includegraphics[width=\textwidth]{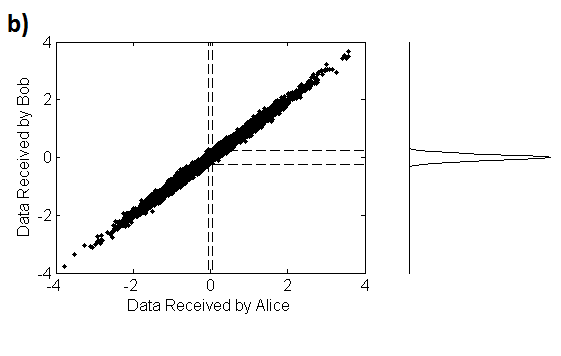}
\end{minipage}%
\begin{minipage}{0.5\textwidth}%
\includegraphics[width=\textwidth]{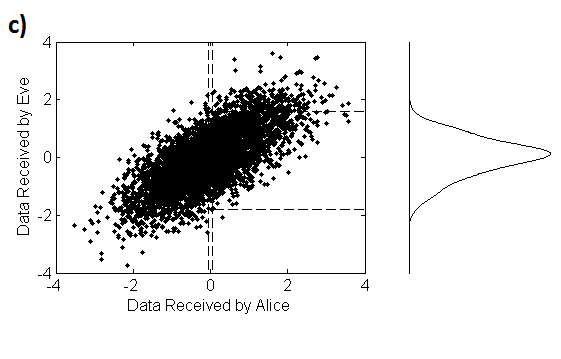}
\end{minipage}%
\caption{A simple experimental realisation of the simulation could look similar to that in {\bf a$)$}. Modulation is applied to a laser beam using an Electro-Optic Modulator (EOM), which is then passed to the detectors of Alice, Bob and Eve. The signal to noise ratio of each detector is controlled by the combination of half wave plates $\left( \frac{\lambda}{2} \right)$ and Polarising Beam Splitters (PBS). Data from simulation at $95\%$ channel transmission is shown in {\bf b$)$} and {\bf c$)$}. Bob's data is very close to that of Alice, while Eve's has considerable noise.}%
\label{fig:3}%
\end{center}
\end{figure}

In order to analyse the slicing methods, a computer was used to simulate a quantum channel, based on that by Grosshans {\it et al}. in \cite{GGnature}. During the simulation, Alice was given a Gaussian distribution of random numbers. Bob and Eve were also given this distribution, but with the addition of some Gaussian noise. The channel transmission was varied so that at high transmissions Bob received few errors, and Eve many; and the reverse at low transmissions. This simulates the transmissions through a real channel and allows us to examine how the channel transmission affects the secrecy. Figure \ref{fig:3} shows the channel and data comparisons for Alice, Bob and Eve. The random number generator was seeded so that each different slicing method always used the same values.

We analysed a range of possible slicing methods which demonstrated our key points. Two properties of the used slicing methods were varied: firstly the size and positioning of the bins, and secondly, the way in which the bins are numbered. A third method, not used here but mentioned for completeness, is the numerical optimisation of the bin positions to give the maximum mutual information between Alice and Bob \cite{oddsl1}. This method is frequently used by Grosshans {\it et al}, who alternate each degree of severity of slicing with an error correction protocol \cite{oddsl2}. As noted above, increasing the mutual information between Alice and Bob does not always increase the secrecy, in fact we show here that sometimes the opposite applies. 

\begin{figure}[pb]
\begin{center}
\includegraphics[width=\linewidth]{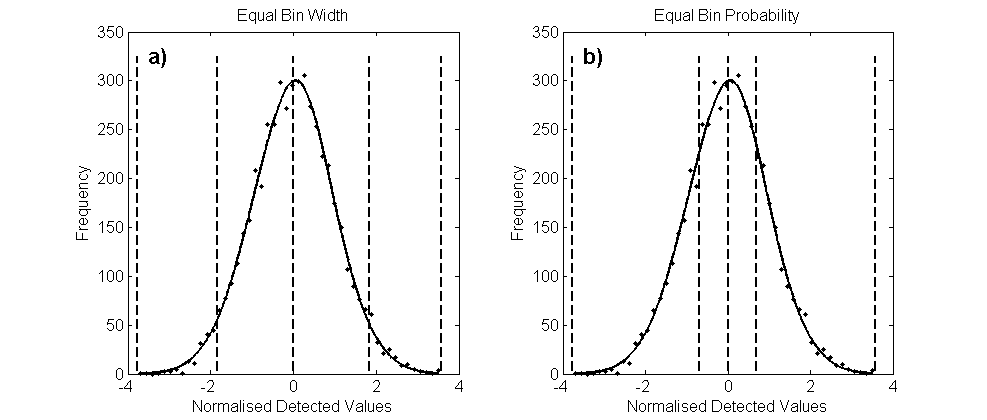}
\caption{The Gaussian distribution of transmitted or received values, as seen by one of Alice, Bob or Eve. The bin edges using {\bf a$)$} equal width, and {\bf b$)$} equal probability for the same curve are marked with dashed lines.}
\label{fig:4}
\end{center}
\end{figure}

Two methods of bin positioning were used during slicing of the Gaussian distribution. In the first, the bins were placed at uniform distances along the x-axis, so that a set range of measured values fell into each bin. In the second method, the bins were chosen so that there were an equal number of transmitted points in each bin. These are demonstrated in figure \ref{fig:4}.

Placing the bins so that they are equal in probability would lead to a higher number of transferred errors than using bins with an equal width. This is due to the bunching of bin boundaries around the centre of the histogram; the centre of the histogram contains more points, and thus more errors, so this placing of bin boundaries will lead to a greater number of transferred errors.

Three methods of bin numbering were used. The first of these was a standard binary code, as this is the simplest and most common method of digital numbering. It is an instinctive choice.

For the second method, a Gray code was chosen as this only has a single (and therefore the minimum) bit difference between adjacent bins. Standard binary on the other hand, frequently has differences of several bits between bins, meaning that an error in transmission which places a point in the bin neighbouring that from its original position will cause more than one bit error in the final string. 

For comparison, a third method was arbitrarily chosen from one of a number of different methods which give significantly more numbers of differing bits between adjacent bins than either binary or Gray, causing the maximum disruption to the acquired string. The method chosen for this case was a Fibonacci Linear Feedback Shift Register (F-LFSR), which works as follows. The first bin is labelled with an appropriate number of `0's' followed by a `1' such as `0001'. For each subsequent bin, an XOR operation is performed between the final two bits, and the resultant bit becomes the first bit of the new label. The other digits are all shifted right by one place, with the one on the far right end being discarded. For example, `0001' becomes `1000' then `0100' and `0010' etc.

For simplicity, in this paper, we limit the eavesdropper to using the same numbering scheme as the legitimate parties. In practice, it may be possible for Eve to gain more information than in this case, increasing her own closeness to the legitimate users.

\begin{figure}[pb]
\begin{center}
\includegraphics[width=.8\linewidth]{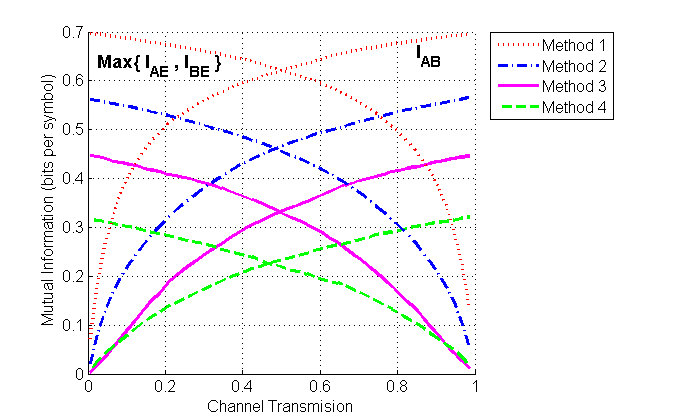}
\caption{(colour online) $I_{AB}$ and $max\{I_{AE} , I_{BE}\}$ against channel transmission for four different slicing methods, each using $2^{4}$ bins. Method 1 represents an F-LFSR code using bins of equal probability, method 2 is F-LFSR with bins of equal width, method 3 a Gray code with bins of equal width and method 4 is a Gray code with bins of equal probability. $Max\{I_{AE}, I_{BE}\}$ is greater than $I_{AB}$ at channel transmissions below about 0.5, meaning that in this region, under these conditions, key cannot be produced.}
\label{fig:5}
\end{center}
\end{figure}

After the data had been sliced, the mutual information between Alice and Bob ($I_{AB}$), Alice and Eve ($I_{AE}$) and Bob and Eve ($I_{BE}$) was calculated, and used to determine $\Delta I$, where $\Delta I = I_{AB} - \max\{ I_{AE} , I_{BE} \}$ \cite{Maurer}. This is a commonly used measure of security, with key production thought to be impossible if $\Delta I$ is non-positive.
\\\\
The results of this model are presented in figures 5-8.
Each of the six different trialled slicing methods were studied, using $2^{4}$, $2^{5}$ and $2^{6}$ bins in each case. Figure \ref{fig:5} shows a comparison of $I_{AB}$ and $\max \{I_{AE} , I_{BE}\}$ against the channel capacity for four different slicing methods. For simplicity, only the numbering methods with the highest (F-LFSR) and lowest (Gray code) numbers of transferred errors are shown, each with both of the different bin positioning methods, and each using $2^{4}$ bins (4 bits to describe each bin).

\begin{figure}[pb]
\begin{center}
\includegraphics[width=\linewidth]{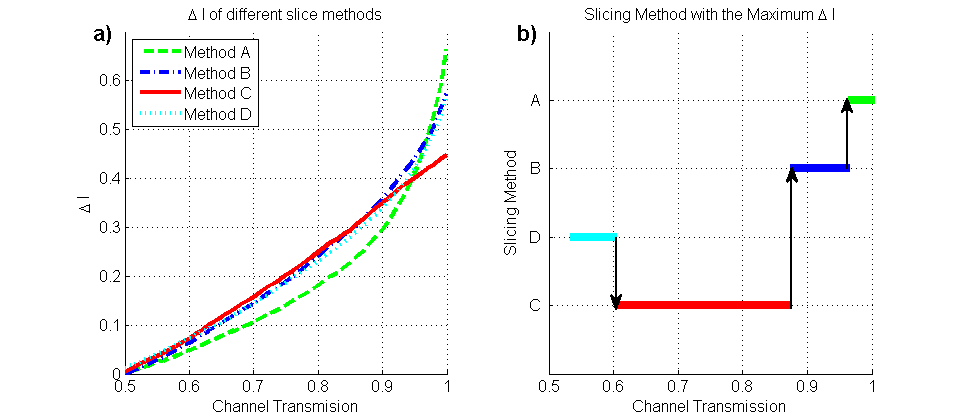}
\caption{(colour online) {\bf a$)$} The secrecy of different slicing methods as a function of channel capacity, and {\bf b$)$} the optimal slicing method for each channel capacity. The slicing methods in {\bf b$)$} are arbitrarily ordered with those having the highest $I_{AB}$, and fewer transferred errors towards the top. Method {\it{A}} uses a Gray code and bins of equal probability, {\it{B}} uses standard binary with bins of equal probability, {\it{C}} is a F-LFSR with bins of equal probability and {\it{D}} is a Gray code with bins of equal width. Generally Alice and Bob favour a method with higher numbers of transferred errors at lower channel transmissions.}
\label{fig:6}
\end{center}
\end{figure}

It can easily be shown that above a channel transmission of 0.5, the information shared between Alice and Bob is higher than the information shared between the eavesdropper and either of the legitimate users. However, below a channel transmission of 0.5, the reverse is true, implying that secrecy is lost, and no key can be made. While it can be possible to reconcile key for channel transmission below 0.5, this is only done using a method known as reverse reconciliation \cite{RRpreprint} which is examined further on in this paper. For this section, only standard direct reconciliation is used.

Figure \ref{fig:6} {\bf a$)$} shows $\Delta I$ against channel transmission for the four most secure slicing methods trialled (those with the highest $\Delta I$). It can be seen that as expected, these four methods have all been sliced with the lowest number of bins, however the numbering system and the position of the bins are not constant throughout these four, with the optimal choice changing with the channel transmission. The optimal slicing method at each channel transmission is shown in figure \ref{fig:6} {\bf b$)$}, with the slicing methods arbitrarily ordered with those towards the top having the highest $I_{AB}$, and thus the lowest number of errors transferred.
 
It can also be seen that at higher transmissions, the codes with fewer transferred errors are preferable in general as Alice and Bob do not want to introduce errors between themselves. However at lower channel transmissions, the codes which transfer more errors are favoured, due to Alice and Bob purposefully introducing errors to try and distance themselves from Eve. There is a slight rise in figure \ref{fig:6}{\bf b$)$} between about $55\%$ and $60\%$ channel transmission. In this region, where $I_{AE}$ is very close to but less than $I_{AB}$, it appears to be advantageous for Alice and Bob to switch to a slicing method with fewer transferred errors to try and increase their information advantage over Eve.

In reverse reconciliation \cite{RRpreprint}, during the classical advantage distillation and error correction protocols, Alice changes her data to match that which Bob has received. Usually, direct reconciliation takes place (Bob changing his data to match what Alice sent), but it has been shown \cite{GGnature} that using reverse reconciliation can enable a secret key to be produced even at less than $50\%$ channel transmission. 

\begin{figure}[pb]
\begin{center}
\includegraphics[width=.8\linewidth]{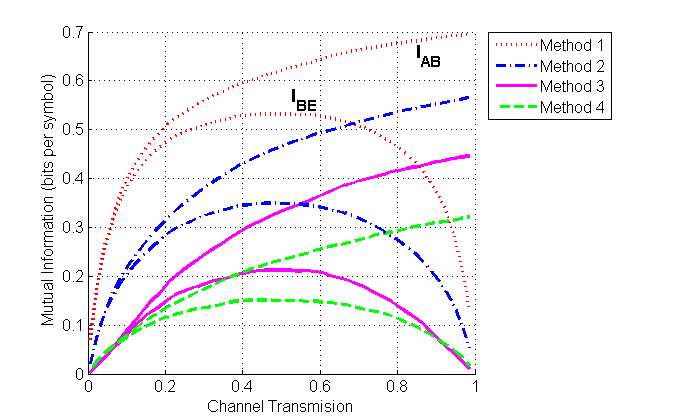}
\caption{(colour online) $I_{AB}$ and $I_{BE}$ against channel transmission for four different slicing methods, each using $2^{4}$ bins. For each slicing method, $I_{BE}$ is consistently below $I_{AB}$. As in fig\ref{fig:5}, method 1 is an F-LFSR with bins of equal probability, method 2 shows an F-LFSR with bins of equal width, method 3 a Gray code with bin of equal probability, method 4 a Grey code with bins of equal width.}
\label{fig:7}
\end{center}
\end{figure}

\begin{figure}[pb]
\begin{center}
\includegraphics[width=\linewidth]{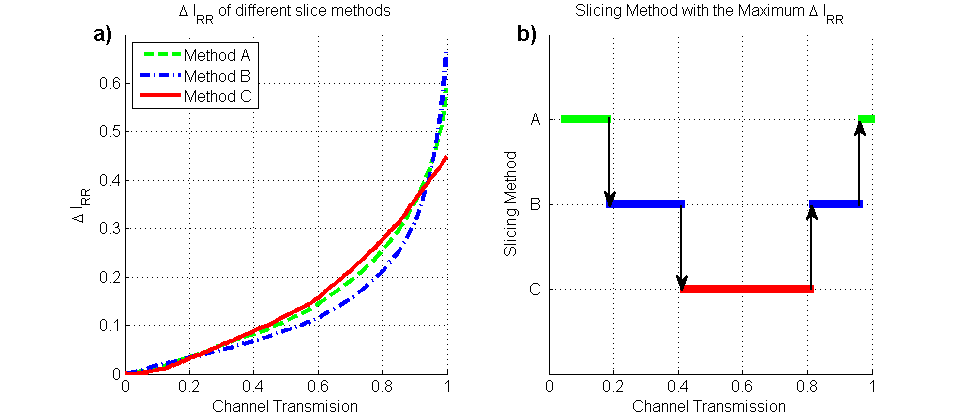}
\caption{(colour online) {\bf a$)$} The secrecy of different slicing methods as a function of channel capacity when using reverse reconciliation, and {\bf b$)$} the optimal slicing methods at each value of channel transmission. As with the direct reconciliation case, {\it A} represents a Gray code with bins of equal probability, {\it B} is standard binary with bins of equal probability, and {\it C} is a F-LFSR also with bins of equal probability. Again, the best slicing method to use changes with the channel transmission.}
\label{fig:8}
\end{center}
\end{figure}

The condition for key distillation in direct reconciliation is $I_{AB} > I_{AE}$, whereas in reverse reconciliation, it is $I_{AB} > I_{BE}$. As can be seen in figure \ref{fig:8}, $I_{AB} > I_{BE}$ for all channel transmissions, so key can always be produced. If $\Delta I_{RR}$ is taken to be $I_{AB} - I_{BE}$, then figure \ref{fig:8} is produced. Figure \ref{fig:8} {\bf a$)$} shows the same trends as when using direct reconciliation and with the same three best slicing methods, but shows $\Delta I$ to be positive for all channel transmissions as expected \cite{GGnature}. Figure \ref{fig:8} {\bf b$)$} also shows the same trend as the direct reconciliation graphs above $50\%$ transmission, but the opposite trend below that. This is because at low channel transmissions, using reverse reconciliation means that while Alice and Bob share very little information, Eve and Bob also share very little information, as is shown in figure \ref{fig:7}. A low error transfer rate will help Alice and Bob, but not necessarily Eve.

\section{Discussion and Conclusions}
\label{sec:3}
Practical QKD systems consist of several different individual protocols, a quantum transmission, followed by at least one classical protocol. The incautious stacking of several of these protocols together can lead to an unexpected lowering of security. In particular, local transformations of data from quantumly received states to classically computable ones during slicing, can unintuitively lose significant amounts of secrecy. 
\\\\
We have shown in figures \ref{fig:6} and \ref{fig:8} that for slicing, the method for which the least secrecy is lost when crossing the quantum classical boundary changes unpredictably with the channel transmission. While general trends are followed, it is impossible to forecast which slicing method is optimal at which values of channel transmission without running prior simulations. Additionally, here only a very few different slicing methods have been trialled, whereas in reality there will be a multitude of different methods which would all need to be examined if the best method for each channel transmission is to be found. 

In general however, for both direct and reverse reconciliation, the smaller the number of bins, the greater the security, and for higher channel transmissions, slicing methods with fewer transferred errors are more optimal. 
\\

\begin{figure}[pb]
\begin{center}
\includegraphics[width=.8\linewidth]{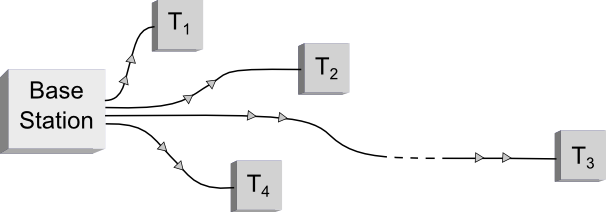}
\caption{An example of a quantum network, where information is sent from the base station to each terminal. The terminals are not necessarily equidistant from the base station, suggesting that each path will have a different channel capacity. This will cause issues in the choice of slicing method when attempting to maximise the secrecy of the network.}
\label{fig:9}
\end{center}
\end{figure}

Both the fact that the best slicing method to choose changes with the channel transmission, and the unpredictability of this have inconvenient consequences for the design of real world CVQKD applications. 

For example, during free space transmissions through the atmosphere (for instance during satellite communications) the channel transmission of the link can vary significantly due to changes in the ionosphere, which can be affected by everything from time of day to solar activity \cite{sat}. These large and frequently unpredictable changes in channel transmission make the choice of slicing method unclear.

Another example would be in a QKD network, an example of which is shown in figure \ref{fig:9}. Here the base station sends out data to separate terminals, all of which are at different distances, and thus are likely to have different channel transmissions. Decisions would then have to be made as to whether all the lines used the same slicing method, or if they should be chosen separately. If they were all the same, the choice of method becomes critical, and if they were different, difficulties in implementation would arise.
\\\\
An important further study would be to not limit Eve to using the same slicing method as Alice and Bob, but letting her choose in each circumstance what is best for her. 

Other factors such as the key rate also need to be considered, as it is possible to increase the key rate at the cost of a higher error rate. An upper bound on an acceptable error rate is likely to be provided by the particular classical error correction codes used subsequently. Many of these however will reduce the key rate at high levels of noise, so a balance needs to be found. The question of whether there exists a slicing method for a particular protocol which does not alter the security of the system is also raised.
\\\\
In addition to this, slicing is only one of a number of proposed reconciliation methods. Others, not examined here, may also be vulnerable to side channel attacks, exploiting either a reduction in security across the quantum classical boundary, or unexpected information leakage in the joining of two or more classical protocols.

\section{Acknowledgements}
E.N., M.E. and F.W. would like to thank the Engineering and Physical Sciences Research Council for their support, F.W. would like to additionally thank Airbus Defence and Space.

\end{document}